# Priority-Aware Near-Optimal Scheduling for Heterogeneous Multi-Core Systems with Specialized Accelerators

Zhuo Chen, and Diana Marculescu, Fellow, IEEE

## ABSTRACT

To deliver high performance in power limited systems, architects have turned to using heterogeneous systems, either CPU+GPU or mixed CPU-hardware systems. However, in systems with different processor types and task affinities, scheduling tasks becomes more challenging than in homogeneous multi-core systems or systems without task affinities. The problem is even more complex when specialized accelerators and task priorities are included. In this paper, we provide a formal proof for the optimal scheduling policy for heterogeneous systems with arbitrary number of resource types, including specialized accelerators, independent of the task arrival rate, task size distribution, and resource processing order. We transform the optimal scheduling policy to a nonlinear integer optimization problem and propose a fast, near-optimal algorithm. An additional heuristic is proposed for the case of priority-aware scheduling. Our experimental results demonstrate that the proposed algorithm is only 0.3% from the optimal and superior to conventional scheduling policies.

## CCS Concepts

• **Software and its engineering** → **Process management**

## Keywords

Heterogeneous systems, performance modeling, queueing theory, optimal scheduling

## 1. INTRODUCTION

Traditionally, performance has been the major goal of system design. Due to Moore's Law [1], processor performance and resulting system throughput enjoyed an exponential increase. However, as Dennard scaling [2] started to break down during the past decade, high power and energy consumption have become the main bottlenecks in system design. To alleviate this problem, computer architects have proposed the concept of heterogeneous systems [3,30], which consist of various types of processors or computing resources. In heterogeneous systems, processors designed with certain task characteristics in mind can deliver both high performance and energy efficiency for those task types. For example, it has been shown [4] that GPUs can perform better than CPUs for image processing tasks and, more generally, for parallel tasks. On the other hand, CPUs are superior to GPUs in processing sequential tasks, like sorting. Other than General-purpose Processors (GP) like CPU and General-Purpose GPU (GPGPU), Specialized Accelerators (SA) can perform even better in a narrower range of applications [5]. For instance, widely available network cards are designed for fast and efficient network communication. In the application space, Convolution Engine [5] was proposed to quickly and efficiently compute convolutions. Therefore, a system consisting of various types of GPs and SAs can accommodate and process different workloads with higher throughput and energy efficiency than traditional homogeneous CPU-only systems. However, new design problems emerge in such complex systems: (1) how do we schedule tasks for maximum system throughput in the presence of multiple GPs and SAs; and (2) how should task scheduling be done when provided with priority requirements of task types.

In this paper, we are directly answering the questions above. We first formally find the optimal scheduling policy of a hetero-system with SAs by using queueing theory. This policy is quite universal since it does not require any special assumptions on the workload, such as Markovian properties for task arrival rate and task size distribution, and can work under any resource processing order. The optimal policy is transformed to an optimization problem which we solve through our proposed heuristic **M**aximize-S**A**-then-G**P** (MAP) to quickly solve for a near-optimal solution. For the case of scheduling with priority requirements, we again formulate it as an optimization problem and solve it with our **MI**nimize-**S**quared-error-independently (MIS) algorithm. Due to lack of real platforms incorporating various SAs, we carried out extensive simulations to demonstrate the effectiveness of MAP and MIS algorithms. Results show that MAP is only 0.3% from the optimal solution and is both faster and more scalable than an existing solver, while MIS can better satisfy the priority requirements by 46% over MAP.

To the best of our knowledge, our work makes the following contributions:

1. We are the first to mathematically determine the optimal scheduling policy for affinity-based hetero-systems with both GPs and SAs.
2. Our optimal scheduling policy is very general and practical, since it makes no assumption on the task arrival rate, task size distribution and resource processing order.
3. We proposed MAP heuristic to quickly solve for near-optimal solutions for systems with arbitrary number of resource types.
4. We are the first to propose and formulate the priority-aware scheduling of affinity-based hetero-systems (including both GPs and SAs) as an optimization problem, and solve it by proposing the MIS heuristic.
5. We extensively simulated our algorithms and demonstrated their effectiveness.

The rest of the paper is organized as follows: section 2 introduces our model of the queueing system. In section 3, we mathematically prove the optimal scheduling policy as well as formulate the priority-aware scheduling. Section 4 proposes MAP and MIS heuristics for solving optimal scheduling and priority-aware scheduling, respectively. Section 5 demonstrates the effectiveness of the algorithms by simulations. Section 6 talks about the related work and section 7 concludes the paper.

• Zhuo Chen and Diana Marculescu are with Carnegie Mellon Univeristy, Pittsburgh, PA 15213.
• E-mail: zhuoc1@andrew.cmu.edu, dianam@cmu.edu.

## 2. PROGRAM MODEL AND CLOSED QUEUEING NETWORK
### 2.1 Program Model

Modern multi-core systems allow the execution of multi-threaded applications which can spawn multiple parallel threads. In this work, we focus on the *task level* and view each thread as a sequence of tasks, as shown in the upper part of Figure 1. Tasks can have various sizes, i.e., different amounts of work, and exhibit diverse characteristics. For example, in a gaming application, one thread may be dedicated to the image processing and another for Artificial Intelligence (AI). In the image processing thread, there can be tasks like edge detection and blurring. In the AI thread, there can be sorting tasks, followed by searching tasks. At any time, each thread has one task running or waiting in one of the $k$ resources. Whenever a task is finished, the next task of its owner thread is sent to one resource based on the scheduling policy.

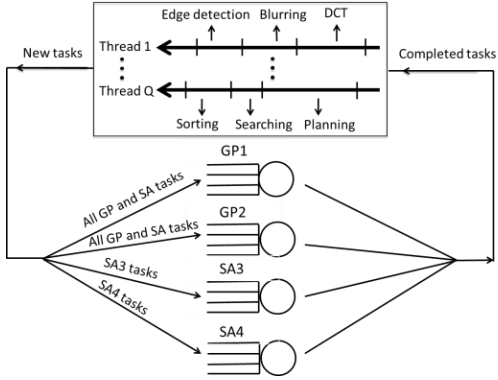

**Figure 1:** $Q$ threads running in an example hetero-system with two GPs and two SAs. Each thread is a sequence of tasks whose characteristics can be different from the others.

In a hetero-system consisting of both General-purpose Processors (GP) and Specialized Accelerators (SA), any given task exhibits different affinities to the GPs or SAs [11]. We say that one task has higher *affinity* to resource A than B, if running on A is better in both performance and energy efficiency than on B [4,31]. In addition, if we have SAs, such as a Convolution Engine (CE), tasks involving convolutions would have highest affinity to CEs rather than any other resource types. Accordingly, we call a task $i$-type if it has the highest affinity to resource type $i$. The existence of affinity property here makes scheduling more challenging than *non-affinity* problems (e.g., in the case of ARM's big.LITTLE [12] and NVIDIA's Tegra [13] architectures), in which big cores are always preferred to small cores for maximizing performance. Instead, this work finds the best policy for scheduling tasks among all available GPs and SAs so as to maximize system performance while considering the task affinity to the resources. Moreover, we satisfy user-defined task type priorities, which measure the different importance of task types.

### 2.2 Closed Queueing Network

In this work, we assume that, within a certain time period, the number of threads $Q$ remains stable. In personal computers, applications are not turned on or off very frequently and this leads to a stable number of running threads, while in the data centers, user threads, which consist of many requests, are relatively stable [14,15]. When the number of threads changes, our policy can be re-calculated on the fly, and in this case, our scheduling policy will be *piece-wise optimal*.

If there are $Q$ threads running, there will always be $Q$ tasks in the system since each thread is a sequence of tasks and whenever a task completes, the following task executes. We can model this system as a *closed batch network* [16]. Figure 1 shows a closed network model of a hetero-system with two GPs and two SAs. GPs can accept any tasks since they are general-purpose, while SAs can only run tasks of its own type since they are specialized.

We define throughput as the rate of task completion $X = \frac{Number\ of\ tasks\ completed}{Elapsed\ time}$, and task response time $T$ as the time from entry to being completed [16]. $T$ is different from the task execution time since includes the waiting time for the resource. In queueing theory, Little's Law [16] for closed batch networks states that:

$$Q = X \cdot \mathbb{E}[T] \quad (1)$$

In other words, since $Q$ is constant, maximizing system throughput is equivalent to minimizing mean task response time. While this does not hold in an open network, our work can be extended to open systems by using the proposed approach in a piece-wise fashion for intervals during which number of threads is constant. Using Little's Law in the context of closed networks is very powerful as it makes no assumptions about task arrival rate, task size distribution, network topology, or resource processing order. This is the main reason for the high applicability of our proposed optimal policy.

## 3. MODELING HETERO-SYSTEMS
### 3.1 System Definition

In this section, we describe the model of general hetero-systems with different types of GPs and SAs. We use indices $1, \ldots, g$ and $g+1, \ldots, g+s = k$ to denote $g$ GPs and $s$ SAs, respectively. Furthermore, since tasks have affinities to certain resource types (e.g., sequential tasks favor CPU and convolutions prefer CE), we define the *affinity matrix* $\mu_{k \times k}$ in which element $\mu_{ij}$ represents the processing rate of $i$-type task on resource $j$. As $j$-type tasks run fastest on resource $j$, we add the affinity constraints: $\mu_{jj} \geq \mu_{ij} \ \forall\ i \neq j$. Furthermore, unlike GPs, SAs can only run tasks of their own type. As a result, $\mu_{ij} = 0$ for (1) $i \leq g$ and $j > g$ (SAs cannot run tasks of GP type), or (2) $i \neq j$ and $i,j > g$ (SAs cannot run tasks of other SA types). The $\mu$ matrix considering all constraints is shown as follows:

$$
\begin{array}{c}
\quad\ \ \ \text{GP1} \quad \cdots \quad \text{GP}g \quad \text{SA}(g+1) \quad \cdots \quad \text{SA}k \\
\begin{array}{c}\text{GP1-task}\\ \cdots \\ \text{GP}g\text{-task}\\ \text{SA}(g+1)\text{-task}\\ \cdots \\ \text{SA}k\text{-task}\end{array}
\left[\begin{array}{cccccc}
\mu_{11} & \cdots & \mu_{1g} & & 0 & \\
\cdots & & \cdots & & & \\
\mu_{g1} & \cdots & \mu_{gg} & & & \\
\mu_{(g+1)1} & \cdots & \mu_{(g+1)g} & \mu_{(g+1)(g+1)} & & 0 \\
\cdots & & & & \cdots & \\
\mu_{k1} & \cdots & \mu_{kg} & 0 & & \mu_{kk}
\end{array}\right]
\end{array} \quad (2)
$$

We further define the task scheduling matrix $N_{k \times k}$, where $N_{ij}$ represents the number of $i$-type tasks within resource $j$'s queue. For similar reasons as above, $N_{ij} = 0$ for (1) $i \leq g$ and $j > g$, or (2) $i \neq j$ and $i,j > g$.

We define the following throughput values based on the definition of throughput: $X_{sys}$ is the entire system throughput; $X_{ij}$ is the throughput of $i$-type task on resource $j$; $X_{*j}$ is the throughput of resource $j$; $X_{i*}$ is the throughput per task type $i$.

Based on these, we can now derive the total system throughput $X_{sys}$ as a function of $\mu_{ij}$ and $N_{ij}$ values. For the processing order for each resource, we use Processor-Sharing (PS), i.e., tasks time-share the resource.[1] Therefore, the time-shared processing rate $\mu_{ij}^*$ for one $i$-type task in resource $j$ is $\mu_{ij}^* = \frac{\mu_{ij}}{\text{Number of tasks in processor } j} = \frac{\mu_{ij}}{\sum_{i=1}^{k} N_{ij}}$. Then, for any GP $j$, $X_{ij} = \mu_{ij}^* \cdot N_{ij} = \frac{\mu_{ij} N_{ij}}{\sum_{i=1}^{k} N_{ij}}$ and $X_{*j} = \sum_{i=1}^{k} X_{ij} = \sum_{i=1}^{k} \frac{\mu_{ij} N_{ij}}{\sum_{i=1}^{k} N_{ij}}$. Therefore, the total throughput from all the GPs is $X_{GP} = \sum_{j=1}^{g} X_{*j} = \sum_{j=1}^{g} \sum_{i=1}^{g} \frac{\mu_{ij} N_{ij}}{\sum_{i=1}^{k} N_{ij}}$. The throughput of one SA is simply $X_{ii} = \mu_{ii}^* \cdot N_{ii} = \mu_{ii}/N_{ii} \cdot N_{ii} = \mu_{ii}$, and the total throughput for all SAs is $X_{SA} = \sum_{i=g+1}^{k} \mu_{ii}$. Finally, the total system throughput is $X_{sys} = X_{GP} + X_{SA} = \sum_{j=1}^{g} \sum_{i=1}^{g} \frac{\mu_{ij} N_{ij}}{\sum_{i=1}^{k} N_{ij}} + \sum_{i=g+1}^{k} \mu_{ii}$. Our goal is determining the optimal scheduling policy that maximizes $X_{sys}$.

### 3.2 Optimal Scheduling Policy

Since the hetero-system is a queueing system, we can solve for the optimal scheduling policy by using a Continuous Time Markov Chain (CTMC) [16]. We define $N_{i*}$ as the total number of $i$-type tasks, i.e., $\sum_{j=1}^{k} N_{ij} = N_{i*}$. Since in a real system, $N_{i*}$ is not determined by system designers but by the running applications, we assume that we already know or have the $N_{i*}$ values. We can see that there are $k - 1$ independent $N_{ij}$'s for each of the $k$ $N_{i*}$ values, and therefore, there are $(k-1) \cdot k$ independent $N_{ij}$'s. With these $N_{ij}$ values, we are able to completely determine the task scheduling matrix $N$. Consequently, we define the system state $S$ as consisting of $(k-1) \cdot k$ independent $N_{ij}$'s: $N_{i_1 j_1}, N_{i_2 j_2}, \ldots, N_{i_{(k-1) \cdot k} j_{(k-1) \cdot k}}$. The system throughput becomes a function of $S$: $X_{sys} = X_{sys}(S)$.

If we assume exponentially distributed task sizes,[2] we can build the CTMC for the hetero-system with all the possible states specified by $S$. The task scheduling probabilities $r(S_i, S_j)$, which determine the scheduling policy, are included in the state transition probabilities of CTMC. The typical way of solving the optimal policy is to calculate the limiting probability $p(S)$ of all the states as a function of all the $r(S_i, S_j)$ values, and then maximize the system throughput $X_{sys} = \sum_{S=1}^{N_S} p(S) \cdot X_{sys}(S)$ with respect to the $r's$. However, instead of solving a large multi-dimensional CTMC problem, we have the following lemma to quickly identify the maximum system throughput, and consequently, better energy efficiency as it's usually the case for affinity-based hetero-systems:

**Lemma 1.** The maximum system throughput is achieved when the system stays in the state $S_{max}$ that can maximize $X_{sys}(S)$. The optimal scheduling policy is to schedule the tasks such that the $N_{ij}$ values can always satisfy $S_{max}$.

*Hint for proof.*[3] $p(S)$ are the CTMC limiting probabilities of each state $S$. We obtain the upper bound ($X_{max} = X_{sys}(S_{max})$) of average system throughput $\bar{X}$ as:

$$\bar{X} = \sum_{S=1}^{N_S} p(S) \cdot X_{sys}(S) \leq \sum_{S=1}^{N_S} p(S) \cdot X_{max} = X_{max} \quad (3) \quad \blacksquare$$

So far, we assumed PS processing order and exponentially distributed task size for illustration purposes. Lemma 2 relaxes these assumptions and shows the generality of our policy.

**Lemma 2.** The optimal scheduling policy is independent of the processing order of the resources and the task size distribution.

*Hint for proof.* With arbitrary task size distributions, we still have $p(S)$ values, though they are not the limiting probabilities any more, and Eq 3 still holds. As the resources process same amount of work when time $\to \infty$, they have same time-average system throughput $\bar{X}$, regardless of the processing order. $\blacksquare$

With lemmas 1 and 2, we showed that the optimal scheduling policy can be found by maximizing the system throughput with respect to state $S$, which consists of $N_{ij}$ values. Therefore, we need to solve the following optimization problem:

$$\text{Maximize} \sum_{j=1}^{g} \sum_{i=1}^{g} \frac{\mu_{ij} N_{ij}}{\sum_{i=1}^{k} N_{ij}} + \sum_{i=g+1}^{k} \mu_{ii} \quad (4)$$

$$S.T. \begin{cases} \sum_{j=1}^{k} N_{ij} = N_{i*}, & i = 1, \ldots, k \\ N_{ij} = 0, & (i \leq g \land j > g) \lor (i \neq j \land i,j > g) \\ N_{ij} \in \mathbb{Z}_{\geq 0}, & i,j = 1, \ldots, k \end{cases} \quad (5)$$

This is a non-linear integer optimization problem with linear constraints which cannot be solved optimally in sub-exponential time. We can exhaustively enumerate all the possible $N_{ij}$ values to find the global optimal, however, this is prohibitive in most systems. To address this issue, we propose an algorithm, **M**aximize-S**A**-then-G**P** (MAP), to quickly solve for a near-optimal solution. MAP is described in detail in section 4.1.

### 3.3 Priority-Aware Scheduling Policy

In some applications, we may want to accelerate certain types of tasks for achieving better performance. With user-defined priorities, we further formulate this priority-aware scheduling policy as an optimization problem. Each task type is assigned a priority value $p_i$ which is user-defined and should match the expected ratio of task type $i$ throughput to total system throughput ($X_{i*}/X_{sys}$). The goal of priority-aware scheduling is to find a schedule for which the resulting $X_{i*}/X_{sys}$ is as close to the user defined $p_i$ as possible. The throughput of $i$-type tasks is $X_{i*} = \sum_{j=1}^{k} X_{ij} = \sum_{j=1}^{k} \frac{\mu_{ij} N_{ij}}{\sum_{i=1}^{k} N_{ij}}$, and the optimization problem is:

$$\text{Minimize} \sum_{i=1}^{k} \left( \frac{X_{i*}}{X_{sys}} - p_i \right)^2 \quad (6)$$

$$S.T. \begin{cases} \sum_{j=1}^{k} N_{ij} = N_{i*}, & i = 1, \ldots, k \\ N_{ij} = 0, & (i \leq g \land j > g) \lor (i \neq j \land i,j > g) \\ N_{ij} \in \mathbb{Z}_{\geq 0}, & i,j = 1, \ldots, k \end{cases} \quad (7)$$

This is also a non-linear integer optimization problem with linear constraints for which we propose a fast heuristic, **MI**nimize-**S**quared-error-independently (MIS). The algorithm is described in detail in section 4.2.

## 4. ALGORITHMS FOR OPTIMIZATIONS
### 4.1 Near-Optimal MAP Algorithm

The MAP algorithm is designed to quickly search for the near-optimal solution of Eq 4 and 5. The pseudocode of MAP is shown in Algorithm 1. It first initializes the $N$ matrix as a diagonal matrix with $N_{ii} = N_{i*}, \forall i$, and then iteratively calls two functions: OptSA (Algorithm 2) which optimizes for SAs, and OptGP (Algorithm 3) which optimizes for GPs, until the solution $N$ converges. The basic idea of OptSA and OptGP is that for each task type, we move the tasks among all the available resources to improve the system throughput, which can be thought as changing the $N_{ij}$ values within each row of matrix $N$. We note that the first term $\sum_{j=1}^{g} \sum_{i=1}^{g} \frac{\mu_{ij} N_{ij}}{\sum_{i=1}^{k} N_{ij}}$ in Eq 4 is actually the sum of $X_{*j}$'s (i.e., GPs' throughput), while the second term is a constant. For each task type $t$, we want to determine which resource will achieve the highest throughput improvement if we move one task to it. For

---

[1] This assumption will be relaxed later.
[2] This assumption will be relaxed later.
[3] Due to space constraints, we are unable to provide full proofs.

this purpose, we can use the sensitivity of $X_{*j}$ with respect to $N_{tj}$ for $j = 1, \ldots, g$:

$$D_{tj} = \frac{\partial X_{*j}}{\partial N_{tj}} = \frac{\partial}{\partial N_{tj}}\left(\sum_{i=1}^{g} \frac{\mu_{ij}N_{ij}}{\sum_{i=1}^{k} N_{ij}}\right) = \frac{\mu_{tj} - X_{*j}}{\sum_{i=1}^{k} N_{ij}} \quad (8)$$

The physical meaning of $D_{tj}$ is as follows: the largest $D_{tj_{max}}$ value indicates that moving $t$-type tasks to resource $j_{max}$, will determine the highest throughput increase. Meanwhile, the smallest $D_{tj_{min}}$ value shows that removing tasks from resource $j_{min}$, will achieve the least throughput penalty. Accordingly, by moving tasks from $j_{max}$ to $j_{min}$, we can achieve the largest throughput improvement. After moving every single task, we need to re-evaluate the $D_{tj}$ values for all $j$'s to keep them up to date.

With the heuristic $D_{tj}$ of Eq 8, OptSA (Algorithm 2) goes through all SA-type tasks (line 4), finds out the $j_{max}$ GP (line 6) and move tasks from SA to it (line 7, 8). OptGP (Algorithm 3) goes through all task types (line 4), finds out $j_{max}$ and $j_{min}$ GPs (line 6) and moves tasks from $j_{min}$ to $j_{max}$ (line 7, 8). Thus, MAP's worst complexity is $O(s \cdot k + k \cdot k) = O(k^2)$.

**Algorithm 1:** Pseudocode of MAP
1: **Input:** $k, g, N_{i*}, \mu$
2: **Output:** $N$
3: $N = diagonal(N_{i*})$
4: **While(1)**
5:     $Ori\_N = N$
6:     $N_{SA} = $ OptSA $(k, g, N, \mu)$
7:     $N_{GP} = $ OptGP $(k, g, N_{SA}, \mu)$
8:     **if** sum(abs($Ori\_N - N$))<1 break
9: **end while**
10: **return** $N$

**Algorithm 2:** Pseudocode of OptSA
1: **Input:** $k, g, N, \mu$
2: **Output:** $N_{SA}$
3: $N_{SA} = N$
4: **for** $t = g + 1: k$ **do**
5:     Calculate $D_{tj}, j = 1, \ldots, g$ [Eq.8]
6:     Find $j_{max}$ that maximizes $D_{tj}$
7:     $N_{SA}[t, t]$ decreases by 1
8:     $N_{SA}[t, j_{max}]$ increases by 1
9: **end for**
10: **return** $N_{SA}$

**Algorithm 3:** Pseudocode of OptGP
1: **Input:** $k, g, N_{SA}, \mu$
2: **Output:** $N_{GP}$
3: $N_{GP} = N_{SA}$
4: **for** $t = 1: k$ **do**
5:     Calculate $D_{tj}, j = 1, \ldots, g$ [Eq.8]
6:     Sort and find $j_{max}$ that maximizes $D_{tj}$ and $j_{min}$ that minimizes $D_{tj}$
7:     $N_{GP}[t, j_{min}]$ decreases by 1
8:     $N_{GP}[t, j_{max}]$ increases by 1
9: **end for**
10: **return** $N_{GP}$

### 4.2 Priority-Aware MIS Algorithm
The basic idea of MIS, shown in Algorithm 4, is to minimize the squared error, $\left(\frac{X_{i*}}{X_{sys}} - p_i\right)^2$ in Eq 6, of each task type $i$ independently. Similar to MAP, for each task type (line 5), we evaluate the $D_{tj}$ values across all the processors (line 7) and move tasks from $j_{min}$ to $j_{max}$ (line 9, 10) until the error does not decrease. Therefore, MIS's worst complexity is $O(k^2)$.

**Algorithm 4:** Pseudocode of MIS
1: **Input:** $k, g, N, \mu$
2: **Output:** $N_{prior}$
3: $N_{prior} = N$
4: **while** $N_{prior}$ not converged
5:     **for** $t$ = priority low to high **do**
6:         **while** squared-error decreases **do**
7:             Calculate $D_{tj}, j = 1, \ldots, k$ [Eq.8]
8:             Sort and find $j_{max}$ that maximizes $D_{tj}$ and $j_{min}$ that minimizes $D_{tj}$
9:             $N_{prior}[t, j_{min}]$ decreases by 1
10:            $N_{prior}[t, j_{max}]$ increases by 1
11:         **end while**
12:     **end for**
13: **end while**
14: **return** $N_{prior}$

## 5. SIMULATION RESULTS

### 5.1 Simulation Configuration
Due to lack of real platforms with various SAs, we simulate a hetero-system in an in-house heterogeneous system simulator implementing our proposed algorithms and several existing ones. As task affinity is expressed via $\mu$ matrix [11], we randomize its entries to show the generality of our methods for widely varying affinities for GPs or SAs [4,30]. To demonstrate that our algorithms can achieve near-optimal solutions regardless of the task size distributions, we use the following commonly used distributions for generating tasks:
1. *Exponential* distribution using Markovian assumption
2. *Bounded Pareto* distribution that many tasks follow [16,19]
3. *Uniform* distribution

We measure the number of tasks completed in a time interval to get the system throughput $X_{sys}$ as well as the throughput per task type $X_{i*}$. The response time of each completed task is recorded and the mean response time of all the tasks $\mathbb{E}[T_{sys}]$ is obtained by averaging over the total task numbers. Furthermore, we use a pessimistic power model in which the power consumption of a task is proportional to its execution time (not response time) in the assigned resource, and report the total simulated energy consumption of the system. The real energy consumption of the system would actually be smaller [4,31]. The corresponding Energy-Delay Product (EDP) is also calculated and reported. We also calculate $X_{sys} \cdot \mathbb{E}[T_{sys}]$ to verify Little's Law.

### 5.2 MAP: Near-Optimal Scheduling with SAs
There are an uncountable number of scheduling policies and we cannot compare with all of them to demonstrate optimality. Instead, we compare MAP against a few widely used polices:
1. Best Fit (BF): $i$-type task is always sent to resource $i$, i.e., $N_{ii} = N_i$.
2. Random (RD): randomly dispatches tasks to the available resources with equal probabilities. A SA-type task will be sent to either one of the GPs or the SA of its type.
3. Load Balancing with perfect information (LB): schedules each task to the resource with the least remaining work so as

to balance the load. This requires that we know exactly how much the remaining load is in each resource. Whenever a SA-type task needs to be scheduled, it tries to balance the load among all the GPs and the SA of its type.

4. Join-the-Shortest-Queue (JSQ): sends each task to the resource with the least number of tasks. When a SA-type task needs to be scheduled, it finds the least-task-number resource among all GPs and the SA of its type.

As MAP is a heuristic, we compare it with an existing solver, Sequential Least SQuares Programming (SLSQP) [20] and the optimal solution found by exhaustive search (Opt). We use SLSQP implemented in the SciPy library, which can efficiently solve relaxed (continuous value) non-linear optimization problem with equality and inequality constraints. Note that SLSQP solves a relaxed optimization problem with continuous value solution instead of integer. Therefore, SLSQP is potentially capable of finding a (unreachable) better solution than the optimal integer one determined by exhaustive search. Because rounding continuous values to integer solution while still keeping the constraints satisfied is not trivial, we keep the continuous value solution, which would be no worse than the rounded integer one, and calculate the corresponding system throughput by Eq 4.

Due to limited space, we only show the simulation results of a hetero-system with two GPs and one SA, e.g., CPU+GPU+CE. $Q = 18$ threads are simulated here, which is determined by $2 * $ size of $N$. We randomize the $\mu$ matrix and $N_{i*}$ values to demonstrate that MAP is effective under different system configurations. $\eta$ on the $x$-axis represents the sample number. We perform 600 simulations and show 20 random samples here, to maintain readability of the figures.

Figures 2, 3 and 4 show that MAP indeed can perform close to the optimal policy (Opt) in terms of throughput, response time and EDP, across all different task size distributions. Averaging over 600 runs, the system throughput delivered by MAP is only 0.3% from the optimal solution. Our simulation results of a two GPs+two SAs system also show that MAP is only 0.5% away from the maximum throughput.

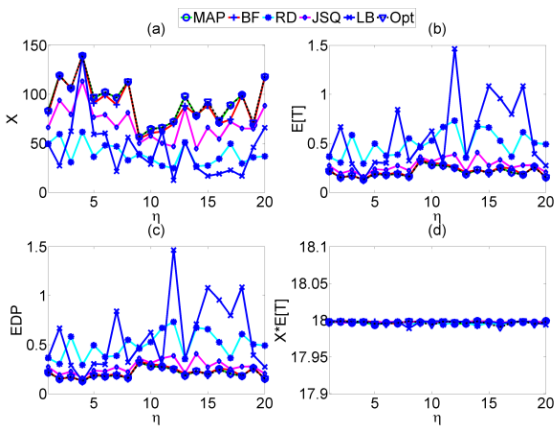

**Figure 2**: Four metrics ($X_{sys}$, $\mathbb{E}[T_{sys}]$, EDP and $X_{sys} \cdot \mathbb{E}[T_{sys}]$) of all six policies under Exponential distribution. MAP achieves similar results as the optimal solution (Opt).

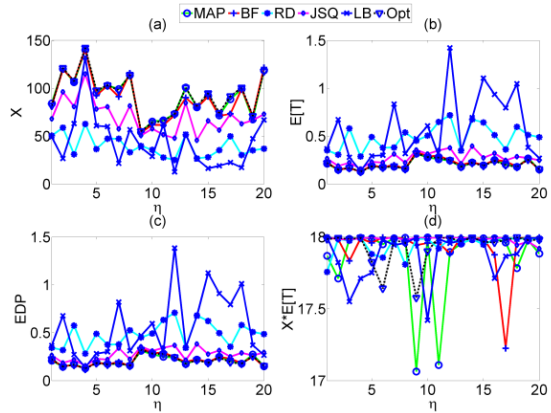

**Figure 3**: Four metrics ($X_{sys}$, $\mathbb{E}[T_{sys}]$, EDP and $X_{sys} \cdot \mathbb{E}[T_{sys}]$) of all six policies under Bounded Pareto distribution. MAP achieves similar results as the optimal solution (Opt).

The $X_{sys} \cdot \mathbb{E}[T_{sys}]$ value is very close to 18, which verifies Little's Law in the closed network. In Bounded Pareto distribution, it actually deviates from 18 more than the other two distributions. This is because Bounded Pareto is known to be heavy-tailed, and this high variation can be reduced by simulating a longer time to ensure enough sampling in the distribution tail [16], which is verified in our simulations.

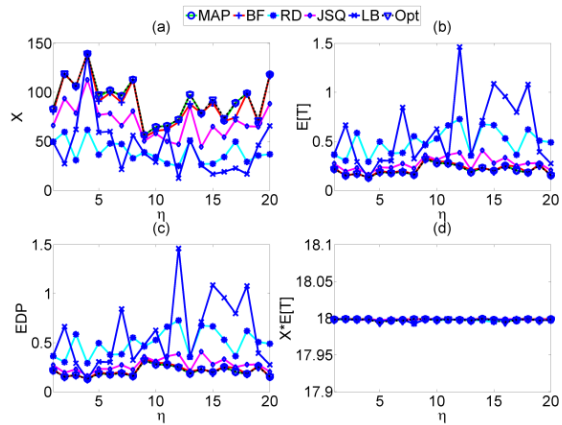

**Figure 4**: Four metrics ($X_{sys}$, $\mathbb{E}[T_{sys}]$, EDP and $X_{sys} \cdot \mathbb{E}[T_{sys}]$) of all six policies under Uniform distribution. MAP achieves similar results as the optimal solution (Opt).

We are also interested in performance of MAP algorithm when we scale up the number of resource types. Since Opt does not scale, we only compare MAP with SLSQP here. Figure 5 shows that MAP with integer solution achieves almost the same throughput values as SLSQP's continuous value solution.

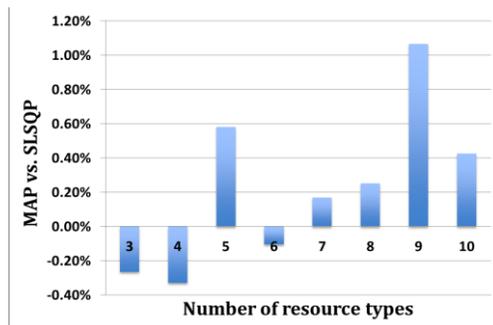

**Figure 5**: Solution comparison of MAP and SLSQP. MAP achieves similar results as SLSQP's continuous value solution.

We further compare the algorithm runtime of MAP and SLSQP. We know that any of these algorithms can quickly terminate, and hence report a shorter runtime, if they find a much worse solution. In an extreme case, either algorithm can report the initial guess as the solution. The consequence is that the runtime is extremely fast but the quality of solution can be low. Accordingly, to make a fair comparison, we only compare the runtime of both algorithms when their throughput values are within 5%. Figure 6 shows the results. We see that MAP is constantly faster than SLSQP. In addition, when there are more resource types, MAP is more scalable than SLSQP as it has a slower increasing rate.

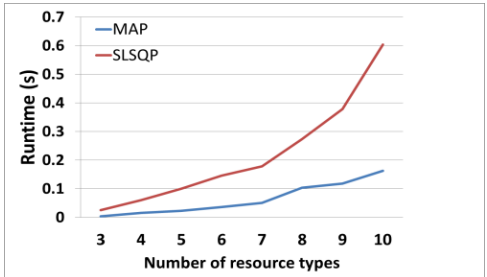

**Figure 6**: Algorithm runtime comparison of MAP and SLSQP. MAP is not only faster but also more scalable than SLSQP.

### 5.3 MIS: Priority-Aware Scheduling

We simulated a hetero-system with MIS and compared it with MAP to show its effectiveness in optimizing for priority requirements. The squared-error improvement is defined as $-\frac{error\ of\ MESI - error\ of\ MAP}{error\ of\ MAP}$. Higher value means that MIS can better minimize squared-error, and hence better satisfy the priority requirement than MAP. Figure 7 demonstrated that MIS really can better satisfy the priority requirements, by 46% on average, than MAP across all different task size distributions. Although Bounded Pareto again shows a higher variation, it can be reduced with longer simulation as pointed out in the previous section.

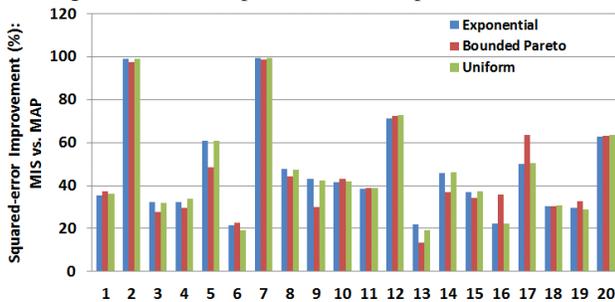

**Figure 7:** Priority improvement of MIS over MAP. On average, MIS better satisfies the priority requirement by 46% than MAP.

## 6. RELATED WORK

Several authors have addressed the problem of performance optimization for hetero-systems [6,7,8,9,10]. Static methods, e.g., [9,10], optimize systems by profiling the programs offline. However, this approach is expensive and cannot accommodate dynamic workload changes. Queueing theory-based approaches [1] attempt to optimize system throughput and power. However, they only work in *non-affinity* problems, and most of them require Markovian assumptions, e.g., Poisson arrival process and exponentially distributed task size, which are often not true, as correctly being pointed out by [19] and [21]. Machine learning methods [7,9,22] have large training cost and cannot guarantee optimality. Other queueing theory-based theoretical results try to solve for the *optimal* hetero-system task scheduling policy [11,23,24,25] but they can only provide approximations via either computational [23] or analytical methods [11]. For example, [26] proposes a myopic policy (assuming no further arrivals). [27] gives asymptotic optimality in heavy traffic regime. Most existing work assume either PS [11] or FCFS [23,28,29] processing order, and that all the resources are GPs without considering SAs. They are tied to Markovian property assumption and thus, are less general. In all, the above approaches rely on restrictive assumptions and can only approximate the optimal task scheduling in all-GP systems, and do not consider task priorities in the hetero-system scheduling.

## 7. CONCLUSION

In this work, we formally determine the optimal scheduling policy for hetero-systems with GPs and SAs and propose the MAP heuristic that can quickly solve for the near-optimal policy, which is only 0.3% from the optimal. By considering the user requirements of the task priorities, we further formulate the priority-aware scheduling as a non-linear integer optimization problem, and propose the MIS heuristic, which can better satisfy the required priorities by 46% over MAP. Extensive simulations demonstrate that our algorithms are not only effective but also very general since they are independent of the task arrival rates, task size distributions and resources' processing orders.